\documentclass[english]{article}
\usepackage[T1]{fontenc}
\usepackage[latin9]{inputenc}
\usepackage{geometry}
\usepackage{float}
\usepackage{amsmath}
\usepackage{stmaryrd}
\usepackage{graphicx}

\makeatletter

\providecommand{\tabularnewline}{\\}

\newcommand{\lyxaddress}[1]{
	\par {\raggedright #1
	\vspace{1.4em}
	\noindent\par}
}

\makeatother

\usepackage{babel}
\begin{document}
\title{Matter Effects on Mass Square Difference for Four Flavor Neutrino
Oscillation}
\author{Vivek Kumar Nautiyal$\mathrm{^{1}}$, Bipin Singh Koranga$\mathrm{^{2}}$,
Ashish Shrivastava$\mathrm{^{2}}$, Neelam Das$\mathrm{^{3}}$}
\maketitle

\lyxaddress{$\mathrm{^{1}}$Department of Physics, Babasaheb Bhimrao Ambedkar
University, Lucknow-226025, India}

\lyxaddress{$\mathrm{^{2}}$Department of Physics, Kirori Mal college (University
of Delhi), Delhi-110007, India}

\lyxaddress{$\mathrm{^{3}}$Department of Physics, University of Lucknow, Lucknow-226007,
India}
\begin{abstract}
\noindent We consider the matter effects on four flavor neutrino oscillation
scheme (3+1). In presence of sterile neutrino, the simplest four flavor
neutrino mixing there are six mixing angles and three Dirac CP phases.
In this paper, we discuss about the sensitivity of mass square difference
effects $\Delta_{21}^{m}$, $\Delta_{31}^{m}$~and $\Delta_{41}^{m}$
in matter. We find that in presence of sterile neutrino for four flavor
mixing framework, only solar mass square difference $\Delta_{21}^{m}$
and atmospheric neutrino mass square difference $\Delta_{31}^{m}$
change for different values of Dirac phases and energy. There is no
change of sterile neutrino mass square differences. In this letter,
we study the matter effects on neutrino mass square differences $\Delta_{21}$,
$\Delta_{31}$~and $\Delta_{41}$~ and calculate the percentage
change with respect to the mass square difference in the matter by
varying energy for four flavor framework.
\end{abstract}

\section{Introduction}

A chargeless and mass particle; neutrino is only interacting via weak
interaction and it does not respect the parity conservation. The mass
of the neutrino has been confirmed by the process called neutrino
oscillation that has been proposed by atmospheric, solar, accelerator,
and reactor neutrino experiments {[}1, 2, 3{]}. The state of neutrino
flavor is related to its mass eigenstate via the PMNS matrix that
contains mixing angles and Dirac phases. The observational results
from reactor neutrino experiment {[}4, 5, 6{]} identified the third
mixing angle, $\theta_{13}$. The pursuit for sterile neutrinos in
experiments has recently become more intense. The Large Hadron Collider
run gives a unique chance for heavy sterile neutrinos. The IceCube
Neutrino Observatory in Antarctica provides a unique perspective.
In addition, an ice cube recently detected a high-energy neutrino
event that is completely outside the framework of SM {[}18, 19{]}.
The results of the LSND and MiniBoone experiments give us a glimpse
into the presence of sterile neutrinos. LSND and MiniBoone experiment
gives solar neutrino mass square difference in the range $0.2<\Delta_{21}<10eV^{2}[7]$
~and~$0.01<\Delta_{21}<1.0eV^{2}${[}8{]}, respectively. These experimental
results provides a new window for the presence of sterile neutrinos.
When the neutrino propagates through the matter, due to charged current
interaction of neutrinos with matter, mass square differences in vacuum
can be modified. The well known MSW effects{[}16{]}, which can be
explain by modified Hamiltonian. The three active neutrinos interacts
by neutral and charged current interaction. But the sterile neutrino
has no any interaction with the matter. When neutrino interacts with
matter, the matter effect change the neutrino mass square differences
and mixing angles. In long baseline neutrino experiments, matter effect
give a significant contribution in neutrino mass square differences
and mixing angles while in short baseline experiment it does not have
such effect {[}17{]}. Since, the probability of oscillation also dependent
on mass square difference and mixing angles, it also get effects due
to matter effect.The article outline is as follows. In Section 2\textbf{,}
we discuss about four flavor neutrino oscillation mixing with sterile
neutrino. In Section 3, we briefly discuss about the matter effects
and effective neutrino mass square difference in four flavor framework.
Section 4, numerical results and the conclusion is given in Section
5.

\section{Four Flavor Neutrino in Vacuum with Sterile Neutrino}

In this section, we consider with four flavor framework (3+1) by assuming
the sterile neutrino of eV range and the mixing of this sterile neutrino
with three different neutrinos. By adding one sterile neutrinos {[}9{]},
there is an increment in mixing angles and CP violating phases in
the PMNS matrix $U_{4\vartimes4}$ which is given by {[}10{]},

\begin{equation}
U=R_{34}(\theta_{34},\delta_{34})R_{24}(\theta_{34})R_{14}(\theta_{14},\delta_{14})R_{23}(\theta_{23})R_{13}(\theta_{13},\delta_{13})R_{12}(\theta_{12}),
\end{equation}

where the matrices $R_{ij}$are rotations in ij space,
\[
R_{ij}(\theta_{ij},\delta)=\left(\begin{array}{cc}
c_{ij} & s_{ij}e^{-i\delta}\\
-s_{ij}e^{i\delta} & c_{ij}
\end{array}\right),
\]

where $s_{ij}=sin\theta_{ij}$,$\,c_{ij}=cos\theta_{ij}$.

Note that in four flavor there are three Dirac CP-violating phase
$\delta_{ij}.$ The explicit form of U is

\begin{equation}
U=\left(\begin{array}{cccc}
U_{e1} & U_{e2} & U_{e3} & U_{e4}\\
U_{\mu1} & U_{\mu2} & U_{\mu3} & U_{\mu4}\\
U_{\tau1} & U_{\tau2} & U_{\tau3} & U_{\tau4}\\
U_{s1} & U_{s2} & U_{s3} & U_{s4}
\end{array}\right),
\end{equation}

and

\[
\mathrm{U=\left(\begin{array}{cccc}
(c_{14}c_{13}c_{12}) & (c_{14}c_{13}s_{12}) & (c_{14}e^{-i\delta_{13}}s_{13}) & s_{14}e^{-i\delta_{13}}\\
\\
(-c_{24}c_{23}s_{12} & (c_{24}c_{23}c_{12} & (-c_{24}s_{23}c_{13} & (s_{24}c_{14})\\
-(c_{24}s_{23}s_{13}c_{12}e^{-i\delta_{13}} & -(c_{24}s_{23}s_{13}s_{12}e^{i\delta_{13}} & -s_{24}s_{14}s_{13}e^{-i(\delta_{13}-\delta_{14})})\\
-s_{24}s_{14}c_{13}c_{12}e^{-i\delta_{14}})) & -s_{24}s_{14}c_{13}s_{12}e^{i\delta_{14}}))\\
\\
(c_{34}s_{23}s_{12} & (-c_{34}s_{23}c_{12} & (c_{34}c_{23}c_{13} & (s_{34}c_{24}c_{14}e^{-i\delta_{34}})\\
-c_{34}c_{23}s_{13}c_{12}e^{i\delta_{13}} & -c_{34}c_{23}s_{13}s_{12}e^{i\delta_{13}} & -s_{34}s_{24}s_{23}c_{13}e^{-i\delta_{34}}\\
+s_{34}s_{24}c_{23}s_{12}e^{-i\delta_{34}} & -s_{34}s_{24}c_{23}c_{12}e^{-i\delta_{34}} & -s_{34}c_{24}s_{14}s_{13}e^{-i(\delta_{13}-\delta_{14}+\delta_{34})})\\
+s_{34}s_{24}s_{23}s_{13}c_{12}e^{-i(\delta_{34}-\delta_{13})} & +s_{34}s_{24}s_{23}s_{13}s_{12}e^{-i(\delta_{34}-\delta_{13})}\\
-s_{34}c_{24}s_{14}c_{13}c_{12}e^{-i(\delta_{34}-\delta_{14})}) & -s_{34}c_{24}s_{14}c_{13}s_{12}e^{-i(\delta_{34}-\delta_{14})})\\
\\
(-s_{34}s_{23}s_{12}e^{i\delta_{34}} & (s_{34}s_{23}c_{12}e^{i\delta_{34}} & (-s_{34}c_{23}c_{13}e^{i\delta_{34}} & (c_{34}c_{24}c_{14})\\
+s_{34}c_{23}s_{13}c_{12}e^{i(\delta_{34}+\delta_{13})} & +s_{34}c_{23}s_{13}s_{12}e^{i(\delta_{34}+\delta_{13})} & -c_{34}s_{24}s_{23}c_{13}\\
+c_{34}s_{24}c_{23}s_{12} & -c_{34}s_{24}c_{23}c_{12} & -c_{34}c_{24}s_{14}s_{13}e^{-i(\delta_{13}-\delta_{14})})\\
+c_{34}s_{24}s_{23}s_{12}c_{12}e^{i\delta_{13}} & +c_{34}s_{24}s_{23}s_{13}s_{12}e^{i\delta_{13}}\\
-c_{34}c_{24}s_{14}c_{13}c_{12}e^{i\delta_{14}}) & -c_{34}c_{24}s_{14}c_{13}s_{12}e^{i\delta_{14}})
\end{array}\right)}
\]

In the presence of one sterile neutrino {[}16{]}, the four flavor
neutrino oscillation probability in vacuum is

\begin{equation}
P_{\nu_{\alpha}\rightarrow\nu_{\beta}}=\delta_{\alpha\beta}-4\sum_{i<j}^{4}Re(U_{\alpha i}U_{\beta j}U_{\alpha j}^{*}U_{\beta i}^{*})sin^{2}\left(\frac{\Delta_{ij}L}{2E}\right)+2\sum_{i<j}^{4}Im(U_{\alpha i}U_{\beta j}U_{\alpha j}^{*}U_{\beta i}^{*})sin2\left(\frac{\Delta_{ij}L}{2E}\right),\,\,\,\,\,\,\,\alpha,\beta=e,\mu,\tau,s
\end{equation}

where L is the baseline length of particular experiment.

\section{Neutrino Mass Square Difference in Matter for Four Flavor}

If we assume the mass eigenstate basis, the Hamiltonian $H_{vacuum}$
in the propagation of neutrinos in vacuum is given by

\begin{equation}
H_{vacuum}=\left(\begin{array}{cccc}
E_{1} & 0 & 0 & 0\\
0 & E_{2} & 0 & 0\\
0 & 0 & E_{3} & 0\\
0 & 0 & 0 & E_{4}
\end{array}\right),
\end{equation}

where $E_{k}(k=1,2,3,4)$ are the energies of the neutrino mass eigenstates
k with mass $m_{k};$

\begin{equation}
E_{k}=\sqrt{m_{k}^{2}+p_{k}^{2}}\approx p_{k}+\frac{m_{k}^{2}}{2p}\approx p+\frac{m_{k}^{2}}{2E}
\end{equation}

we consider the momentum p is the same for all mass eigenstates. When
neutrino interact with matter by weak interaction (charged and neutral
current). Sterile neutrino itself not take any participitation of
weak interaction. The effective Hamiltonian for four flavor neutrino
mixing is {[}10{]}

\begin{equation}
H_{eff}=\frac{1}{2E}\left[U\left(\begin{array}{cccc}
m_{1}^{2} & 0 & 0 & 0\\
0 & m_{2}^{2} & 0 & 0\\
0 & 0 & m_{3}^{2} & 0\\
0 & 0 & 0 & m_{4}^{2}
\end{array}\right)U^{\dagger}+\left(\begin{array}{cccc}
A & 0 & 0 & 0\\
0 & 0 & 0 & 0\\
0 & 0 & 0 & 0\\
0 & 0 & 0 & A^{'}
\end{array}\right)\right],
\end{equation}

where U is four flavor mixing matrix and $A\,\,\,and\,\,\,\,A^{'}$is
matter dependent term is given by

\[
A(eV^{2})=2\sqrt{2}G_{F}N_{e}E_{\nu},
\]

\[
A^{'}(eV^{2})=-\sqrt{2}G_{F}N_{n}E_{\nu},
\]

where $N_{e}$ and $N_{n}$ is the density of electron and neutron.
From Eq.(6), we have

\begin{equation}
\left[U\left(\begin{array}{cccc}
m_{1}^{2} & 0 & 0 & 0\\
0 & m_{2}^{2} & 0 & 0\\
0 & 0 & m_{3^{2}} & 0\\
0 & 0 & 0 & m_{4}^{2}
\end{array}\right)U^{\dagger}\right]=U_{m}=\left[U\left(\begin{array}{cccc}
0 & 0 & 0 & 0\\
0 & \Delta_{21} & 0 & 0\\
0 & 0 & \Delta_{31} & 0\\
0 & 0 & 0 & \Delta_{41}
\end{array}\right)U^{\dagger}\right]+\left[\left(\begin{array}{cccc}
A & 0 & 0 & 0\\
0 & 0 & 0 & 0\\
0 & 0 & 0 & 0\\
0 & 0 & 0 & A^{'}
\end{array}\right)\right]
\end{equation}

where 

\[
\Delta_{ij}=m_{i}^{2}-m_{j}^{2}.
\]

\begin{equation}
a_{4}\lambda^{4}+a_{3}\lambda^{3}+a_{2}\lambda^{2}+a_{1}\lambda+a_{0}=0
\end{equation}

where the coefficients $a_{4},\,a_{3},\,a_{2},\,a_{1}\,and\,a_{0}\,$are
given as

\[
\begin{array}{ccc}
a_{4}=1 & , & a_{3}=-\sum_{\alpha}\sum_{i}\Delta_{i1}|U_{\alpha i}|^{2}-A-A',\end{array}
\]

\[
a_{2}=AA^{'}+A^{'}\sum_{\alpha\neq s}\sum_{i}\Delta_{i1}|U_{\alpha i}|^{2}+A\sum_{\alpha\neq e}\sum_{i}\Delta_{i1}|U_{\alpha i}|^{2}+\sum_{i<j}\Delta_{i1}\Delta_{j1}\sum_{\alpha<\beta}\left(|U_{\alpha i}U_{\beta j}-U_{\alpha j}U_{\beta i}|\right)^{2},
\]
\[
\begin{array}{c}
a_{1}=-AA^{'}\sum_{i}\Delta_{i1}(|U_{\mu i}|^{2}+|U_{\tau i}|^{2})-\sum_{i<j}\Delta_{i1}\Delta_{j1}\left\{ A^{'}\left(\sum_{\alpha<\beta;\neq s}|U_{\alpha j}U_{\beta i}-U_{\alpha i}U_{\beta j}|\right)^{2}+A\left(\sum_{\alpha<\beta;\neq e}|U_{\alpha j}U_{\beta i}-U_{\alpha i}U_{\beta j}|\right)^{2}\right\} \\
-\Delta_{21}\Delta_{31}\Delta_{41}\left\{ \sum_{\alpha<\beta<\gamma}\sum_{i,j,k}\epsilon_{ijk}^{2}\left(|U_{\alpha i}U_{\beta j}U_{\gamma k}-U_{\alpha i}U_{\beta k}U_{\gamma j}|\right)^{2}-2\sum_{\alpha<\beta<\gamma}\sum_{i,j,k}\epsilon_{ijk}^{2}U_{\alpha i}U_{\alpha j}U_{\beta j}U_{\beta k}U_{\gamma i}U_{\gamma k}\right\} 
\end{array}
\]

\[
\begin{array}{c}
a_{0}=AA^{'}\sum_{i<j}\Delta_{i1}\Delta_{j1}\left(|U_{\mu j}U_{\tau i}-U_{\mu i}U_{\tau j}|\right)^{2}+\Delta_{21}\Delta_{31}\Delta_{41}\left[A^{'}\left(\sum_{i,j,k}\epsilon_{ijk}U_{ei}U_{\mu j}U_{\tau k}\right)^{2}+A\left(\sum_{i,j,k}\epsilon_{ijk}U_{\mu i}U_{\tau j}U_{sk}\right)^{2}\right]\end{array}
\]
where, $\,\epsilon_{ijk}\,$ is Levi-Civita symbol and $i,j=2,3,4\,;\,\alpha,\beta,\gamma=e,\mu,\tau,s$
\[
\begin{array}{ccccc}
P=\frac{3a_{3}^{2}-8a_{2}a_{4}}{12a_{4}^{2}} &  & , &  & X=-a_{3}^{3}+4a_{3}a_{2}a_{4}-8a_{1}a_{4}^{2},\end{array}
\]

\[
\begin{array}{ccccc}
L=a_{2}^{2}-3a_{3}a_{1}+12a_{0}a_{4} &  & , &  & M=2a_{2}^{3}-9a_{3}a_{1}a_{2}+27a_{3}^{2}a_{0}+27a_{1}^{2}a_{4}-72a_{2}a_{0}a_{4},\end{array}
\]

\[
Q=\sqrt{P+\frac{1}{3a_{4}}\left[\left(\frac{M}{2}+\sqrt{\left(\frac{M}{2}\right)^{2}-L^{3}}\right)^{1/3}+L\left(\frac{M}{2}+\sqrt{\left(\frac{M}{2}\right)^{2}-L^{3}}\right)^{-1/3}\right]}
\]

The effective mass square difference in matter can be calculated by
the diagonalization of the matrix $U_{m}${[}17{]}. The roots of Eq.(7.0),
gives matter dependent mass square $m_{m1}^{2},m_{m2}^{2},m_{m3}^{2},m_{m4}^{2}$

\[
\lambda_{1}=m_{m1}^{2}=-\frac{a_{3}}{4}-\frac{Q}{2}-\frac{1}{2}\sqrt{3P-Q^{2}-\frac{X}{4Q}},
\]

\[
\lambda_{2}=m_{m2}^{2}=-\frac{a_{3}}{4}-\frac{Q}{2}+\frac{1}{2}\sqrt{3P-Q^{2}-\frac{X}{4Q}},
\]
\[
\lambda_{3}=m_{m3}^{2}=-\frac{a_{3}}{4}+\frac{Q}{2}-\frac{1}{2}\sqrt{3P-Q^{2}+\frac{X}{4Q}},
\]

\[
\lambda_{4}=m_{m4}^{2}=-\frac{a_{3}}{4}+\frac{Q}{2}+\frac{1}{2}\sqrt{3P-Q^{2}+\frac{X}{4Q}}.
\]

Using matter dependent mass square $m_{m1}^{2},m_{m2}^{2},m_{m3}^{2},m_{m4}^{2}$,we
can write matter dependent mass square difference for four flavor
neutrino oscillation

\begin{equation}
\Delta_{21}^{m}=m_{m2}^{2}-m_{m1}^{2}=\sqrt{3P-Q^{2}-\frac{X}{4Q}},
\end{equation}

\begin{equation}
\Delta_{31}^{m}=m_{m3}^{2}-m_{m1}^{2}=Q+\frac{1}{2}\left(\sqrt{3P-Q^{2}-\frac{X}{4Q}}-\sqrt{3P-Q^{2}+\frac{X}{4Q}}\right),
\end{equation}

\begin{equation}
\Delta_{41}^{m}=m_{m4}^{2}-m_{m1}^{2}=Q+\frac{1}{2}\left(\sqrt{3P-Q^{2}-\frac{X}{4Q}}+\sqrt{3P-Q^{2}+\frac{X}{4Q}}\right).
\end{equation}

P, Q and X are depend on neutrino oscillation parameter and density
{[}10{]}.

\noindent Percentage change in mass square differences due to matter
effect with respect to the matter are given as:

\begin{equation}
\varDelta_{21}^{m}\%change=\frac{|\varDelta_{21}-\varDelta_{21}^{m}|}{\varDelta_{21}^{m}}\times100
\end{equation}

\begin{equation}
\varDelta_{31}^{m}\%change=\frac{|\varDelta_{31}-\varDelta_{31}^{m}|}{\varDelta_{31}^{m}}\times100
\end{equation}

\begin{equation}
\varDelta_{41}^{m}\%change=\frac{|\varDelta_{41}-\varDelta_{41}^{m}|}{\varDelta_{41}^{m}}\times100
\end{equation}

\section{Numerical Analysis}

Equation (9.0) to eq.(11), for modified neutrino mass square difference
in the matter for four flavor and from eq. (12) to eq.(14), the percentage
change in mass square difference are calculated with respect to matter.
All mass square differences depend on the type of neutrino mass order.
In this calculation, we assume for normal mass order. We assume matter
density for electron $\rho_{e}=3g/cm^{3}$ and for neutron $\rho_{n}=3g/cm^{3}$,
respectively. Due to matter effects, the modifed mass square difference
in matter~$\Delta_{21}^{m},\Delta_{31}^{m}$ and $\Delta_{41}^{m}$
are depend on Dirac phases and 6 neutrino mixing angle and matter
density. In calculation, we choose mixing angle $\theta_{12}=34^{o},\,\theta_{23}=45^{o}$
,$\theta_{13}=10^{o}.$ and drop out majorana and charge lepton phases$.$
In this work, we consider following value for sterile neutrino mixing
angles {[}11{]}, $\theta_{14}=3.6^{o},\theta_{24}=4^{o},\theta_{34}=18.5^{o}.\,$We
have taken $\Delta_{31}=0.002eV^{2}${[}12{]},$\Delta_{41}=1.7eV^{2}[11]$
and $\Delta_{21}=0.00008eV^{2}${[}13{]}. In table (1.0) to table
(5.0), we list the modified maximum neutrino mass square difference
in matter for four flavor framework for 1 GeV, 2 GeV and 3 GeV region
for different value of Dirac phases $0^{o}<\delta_{34}<180^{o},$
$\mathrm{0^{o}<\delta_{13}<180^{o},0^{o}<\delta_{14}<180^{o}}$. In
the graph, percentage change in mass square differnces of neutrino
is plotted for energy range 0.1 GeV to 10 GeV ($\varDelta_{i1}^{m}\%$
v/s $Energy$ ,where $i=2,3,4$).

\begin{table}[H]
\begin{tabular}{|c|c|c|}
\hline 
$\Delta_{ij}^{m}$ & Energy (GeV) & $\Delta_{ij}^{m}$\% Change (Maximum)\tabularnewline
\hline 
\hline 
$\Delta_{21}^{m}$ & 1 & 61.33\tabularnewline
\hline 
$\Delta_{31}^{m}$ & 1 & 2.33\tabularnewline
\hline 
$\Delta_{41}^{m}$ & 1 & 0.00\tabularnewline
\hline 
$\Delta_{21}^{m}$ & 2 & 80.97\tabularnewline
\hline 
$\Delta_{31}^{m}$ & 2 & 2.27\tabularnewline
\hline 
$\Delta_{41}^{m}$ & 2 & 0.00\tabularnewline
\hline 
$\Delta_{21}^{m}$ & 3 & 87.42\tabularnewline
\hline 
$\Delta_{31}^{m}$ & 3 & 1.90\tabularnewline
\hline 
$\Delta_{41}^{m}$ & 3 & 0.00\tabularnewline
\hline 
\end{tabular}\caption{For $\delta_{34}=0^{o}$, maximum percentage change in mass square
difference due to matter effect. We have taken $\Delta_{31}=2.0\times10^{-3}eV^{2}$,
$\Delta_{21}=8.0\times10^{-5}eV^{2}$, $\Delta_{41}=1.7eV^{2}$ and
mixing angles $\theta_{13}=10^{o},\theta_{23}=45^{o},\theta_{12}=34^{o}$
,$\theta_{34}=18.5^{o},\theta_{24}=4^{o},\theta_{14}=3.6^{o}$}
\end{table}

\begin{table}[H]
\begin{tabular}{|c|c|c|}
\hline 
$\Delta_{ij}^{m}$ & Energy (GeV) & $\Delta_{ij}^{m}$\% Change (Maximum)\tabularnewline
\hline 
\hline 
$\Delta_{21}^{m}$ & 1 & 61.37\tabularnewline
\hline 
$\Delta_{31}^{m}$ & 1 & 2.25\tabularnewline
\hline 
$\Delta_{41}^{m}$ & 1 & 0.00\tabularnewline
\hline 
$\Delta_{21}^{m}$ & 2 & 81.00\tabularnewline
\hline 
$\Delta_{31}^{m}$ & 2 & 2.11\tabularnewline
\hline 
$\Delta_{41}^{m}$ & 2 & 0.00\tabularnewline
\hline 
$\Delta_{21}^{m}$ & 3 & 87.44\tabularnewline
\hline 
$\Delta_{31}^{m}$ & 3 & 1.66\tabularnewline
\hline 
$\Delta_{41}^{m}$ & 3 & 0.00\tabularnewline
\hline 
\end{tabular}\caption{For $\delta_{34}=45^{o}$, maximum percentage change in mass square
difference due to matter effect. We have taken $\Delta_{31}=2.0\times10^{-3}eV^{2}$,
$\Delta_{21}=8.0\times10^{-5}eV^{2}$, $\Delta_{41}=1.7eV^{2}$ and
mixing angles $\theta_{13}=10^{o},\theta_{23}=45^{o},\theta_{12}=34^{o}$
,$\theta_{34}=18.5^{o},\theta_{24}=4^{o},\theta_{14}=3.6^{o}$}
\end{table}

\begin{table}[H]
\begin{tabular}{|c|c|c|}
\hline 
$\Delta_{ij}^{m}$ & Energy (GeV) & $\Delta_{ij}^{m}$\% Change (Maximum)\tabularnewline
\hline 
\hline 
$\Delta_{21}^{m}$ & 1 & 61.47\tabularnewline
\hline 
$\Delta_{31}^{m}$ & 1 & 2.08\tabularnewline
\hline 
$\Delta_{41}^{m}$ & 1 & 0.00\tabularnewline
\hline 
$\Delta_{21}^{m}$ & 2 & 81.09\tabularnewline
\hline 
$\Delta_{31}^{m}$ & 2 & 1.74\tabularnewline
\hline 
$\Delta_{41}^{m}$ & 2 & 0.00\tabularnewline
\hline 
$\Delta_{21}^{m}$ & 3 & 87.50\tabularnewline
\hline 
$\Delta_{31}^{m}$ & 3 & 1.09\tabularnewline
\hline 
$\Delta_{41}^{m}$ & 3 & 0.00\tabularnewline
\hline 
\end{tabular}\caption{For $\delta_{34}=90^{o}$, maximum percentage change in mass square
difference due to matter effect. We have taken $\Delta_{31}=2.0\times10^{-3}eV^{2}$,
$\Delta_{21}=8.0\times10^{-5}eV^{2}$, $\Delta_{41}=1.7eV^{2}$ and
mixing angles $\theta_{13}=10^{o},\theta_{23}=45^{o},\theta_{12}=34^{o}$
,$\theta_{34}=18.5^{o},\theta_{24}=4^{o},\theta_{14}=3.6^{o}$}
\end{table}

\begin{table}[H]
\begin{tabular}{|c|c|c|}
\hline 
$\Delta_{ij}^{m}$ & Energy (GeV) & $\Delta_{ij}^{m}$\% Change (Maximum)\tabularnewline
\hline 
\hline 
$\Delta_{21}^{m}$ & 1 & 61.57\tabularnewline
\hline 
$\Delta_{31}^{m}$ & 1 & 1.91\tabularnewline
\hline 
$\Delta_{41}^{m}$ & 1 & 0.00\tabularnewline
\hline 
$\Delta_{21}^{m}$ & 2 & 81.18\tabularnewline
\hline 
$\Delta_{31}^{m}$ & 2 & 1.37\tabularnewline
\hline 
$\Delta_{41}^{m}$ & 2 & 0.00\tabularnewline
\hline 
$\Delta_{21}^{m}$ & 3 & 87.56\tabularnewline
\hline 
$\Delta_{31}^{m}$ & 3 & 0.53\tabularnewline
\hline 
$\Delta_{41}^{m}$ & 3 & 0.00\tabularnewline
\hline 
\end{tabular}\caption{For $\delta_{34}=135^{o}$, maximum percentage change in mass square
difference due to matter effect. We have taken $\Delta_{31}=2.0\times10^{-3}eV^{2}$,
$\Delta_{21}=8.0\times10^{-5}eV^{2}$, $\Delta_{41}=1.7eV^{2}$ and
mixing angles $\theta_{13}=10^{o},\theta_{23}=45^{o},\theta_{12}=34^{o}$
,$\theta_{34}=18.5^{o},\theta_{24}=4^{o},\theta_{14}=3.6^{o}$}
\end{table}

\begin{table}[H]
\begin{tabular}{|c|c|c|}
\hline 
$\Delta_{ij}^{m}$ & Energy (GeV) & $\Delta_{ij}^{m}$\% Change (Maximum)\tabularnewline
\hline 
\hline 
$\Delta_{21}^{m}$ & 1 & 61.61\tabularnewline
\hline 
$\Delta_{31}^{m}$ & 1 & 1.83\tabularnewline
\hline 
$\Delta_{41}^{m}$ & 1 & 0.00\tabularnewline
\hline 
$\Delta_{21}^{m}$ & 2 & 81.21\tabularnewline
\hline 
$\Delta_{31}^{m}$ & 2 & 1.21\tabularnewline
\hline 
$\Delta_{41}^{m}$ & 2 & 0.00\tabularnewline
\hline 
$\Delta_{21}^{m}$ & 3 & 87.59\tabularnewline
\hline 
$\Delta_{31}^{m}$ & 3 & 0.30\tabularnewline
\hline 
$\Delta_{41}^{m}$ & 3 & 0.00\tabularnewline
\hline 
\end{tabular}\caption{For $\delta_{34}=180^{o}$, maximum percentage change in mass square
difference due to matter effect. We have taken $\Delta_{31}=2.0\times10^{-3}eV^{2}$,
$\Delta_{21}=8.0\times10^{-5}eV^{2}$, $\Delta_{41}=1.7eV^{2}$ and
mixing angles $\theta_{13}=10^{o},\theta_{23}=45^{o},\theta_{12}=34^{o}$
,$\theta_{34}=18.5^{o},\theta_{24}=4^{o},\theta_{14}=3.6^{o}$}
\end{table}

\begin{figure}[H]
\includegraphics[scale=0.8]{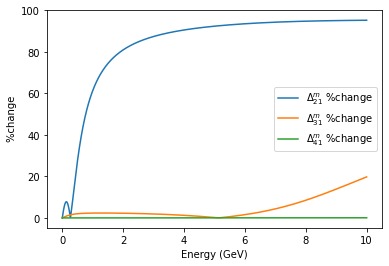}

\caption{Graph $\varDelta_{i1}^{m}\%$ v/s $Energy$ where $i=2,3,4$ We have
taken $\Delta_{31}=2.0\times10^{-3}eV^{2}$, $\Delta_{21}=8.0\times10^{-5}eV^{2}$,
$\Delta_{41}=1.7eV^{2}$ and mixing angles $\theta_{13}=10^{o},\theta_{23}=45^{o},\theta_{12}=34^{o}$
,$\theta_{34}=18.5^{o},\theta_{24}=4^{o},\theta_{14}=3.6^{o}$ and
consider $\delta_{34}=0^{o},\delta_{13}=0^{o}\,and\,\delta_{14}=0^{o}$
.}
\end{figure}

\section{Conclusions}

In this paper, we discussed matter effects on three mass square differences
in four flavor framework. We calculate the modified neutrino mass
square difference for the four flavor mixing model due to matter effects.
Based on our numerical analysis, due to matter effects, this model
predicts a value for the percentage in mass square difference with
respect to the matter $\Delta_{21}^{m}\%change\approx61\%,81\%,87\%$
for 1, 2, 3 GeV respectively. $\Delta_{31}^{m}\%change$ is between
0.3\% to 2.3\% for the region 1 to 3 GeV, and there is no any significant
change of $\Delta_{41}^{m}$sterile neutrino mass square difference.
In the graph, $\varDelta_{21}^{m}\%\,$change behaved differently
in region from 0.0 GeV to approx 0.5 GeV and then it increased upto
more than 80\% as energy increased more than 2\% and the change become
stable in the range of 90\% change after 5 GeV of energy. $\Delta_{31}^{m}\%change$
is not effective as it is in $\Delta_{21}^{m}$. It's plot shows that
the change with respect to the matter first increase upto 2\% and
then decreased about 5 GeV of energy. After 5 GeV it increased upto
20\% from 5 Gev to 10.0 GeV energy.$\Delta_{41}^{m}\%change$ is with
0 slope here, hence there is no any significant change in $\Delta_{41}$due
to matter effect. In conclusion, due to large change of solar mass
square $\Delta_{21}^{m}$ difference in matter, this indicates in
four flavor solar mass square gives large contribution in neutrino
oscillation probability.

\end{document}